\documentclass[letterpaper]{natureprintstyle}
\usepackage{graphicx,epsfig}
\usepackage{amsmath,amssymb,bbm}
\usepackage{color}
\usepackage[usenames,dvipsnames]{xcolor}
\usepackage[colorlinks=true,linkcolor=Red,citecolor=Green,linktoc=page]{hyperref}
\usepackage{multirow}
\usepackage[varg]{txfonts}
\usepackage{ulem}
\usepackage{fancyhdr}

\usepackage[caption=false]{subfig}
\usepackage{balance}
\usepackage{float}
\usepackage{flushend}
\usepackage{graphicx}
\usepackage{amsmath}
\usepackage{upgreek}
\usepackage{colortbl}

\DeclareFontFamily{U}{rcjhbltx}{}
\DeclareFontShape{U}{rcjhbltx}{m}{n}{<->rcjhbltx}{}
\DeclareSymbolFont{hebrewletters}{U}{rcjhbltx}{m}{n}

\usepackage[misc]{ifsym}

\usepackage[caption=false]{subfig}
\DeclareCaptionListOfFormat{nparens}{#1#2}
\usepackage[varg]{txfonts}
\usepackage{ulem}
\usepackage{fancyhdr}

\newcommand{\addresses}[1]{
\thispagestyle{fancy} \lfoot{\parbox{\textwidth}{ \vspace{0.3cm}
 \rule{\textwidth}{0.2pt}
\hspace{-0.2cm} \textsf{\scalefont{0.80} #1} \vspace{-0.2cm}
\begin{center}{\scalefont{0.87} \thepage}\end{center}}}
\cfoot{} }

\DeclareMathSymbol{\lamed}{\mathord}{hebrewletters}{108}

\usepackage[switch]{lineno}

\begin{document}



\title{Microscopic structure of the vortex cores in granular niobium: A coherent quantum puzzle}


\author{V.\,S.\,Stolyarov$^{1,2}$${\textrm{\Letter}}$, 
        V.\,Neverov,$^{2}$
        A.\,V.\,Krasavin,$^{2,3}$
        D.\,I.\,Kasatonov,$^{2}$
        D.\,Panov,$^{2}$
        D.\,Baranov,$^{2}$
	O.\,V.\,Skryabina,$^{2}$
	A.\,S.\,Mel'nikov,$^{2,4,5}$
	A.\,A.\,Golubov,$^{2}$
	M.\,Yu.\,Kupriyanov,$^{2,6}$
        A.\,A\, Shanenko,$^{2,3}$
	T.\,Cren,$^{7}$
	A.\,Yu.\,Aladyshkin,$^{2,4,5}$
        A.\,Vagov,$^{2,3}$
	\& D.\,Roditchev$^{1}$ 
	}
\maketitle

\addresses{
$^1$LPEM, UMR-8213, ESPCI Paris, PSL, CNRS, Sorbonne University, 75005 Paris, France;
$^2$Moscow Institute of Physics and Technology, 141700 Dolgoprudny, Russia;
$^3$HSE University, 101000 Moscow, Russia;
$^4$Lobachevsky State University of Nizhny Novgorod, 603022 Nizhny Novgorod, Russia;
$^5$Institute for Physics of Microstructures RAS, 603950 Nizhny Novgorod, Russia;
$^6$Skobeltsyn Institute of Nuclear Physics, Lomonosov Moscow State University;
$^7$Institut des Nanosciences de Paris, UMR-7588, Sorbonne University, CNRS, F-75252 Paris, France
${\textrm{\Letter}}$ e-mail:stolyarov.vs@phystech.edu
}

\medskip

\textbf{When macroscopic quantum condensates -- superconductors, superfluids, cold atoms and ions, polaritons etc. -- are put in rotation, a quantum vortex lattice forms inside. In  homogeneous type-II superconductors, each vortex has a tiny core where the superconducting gap $\Delta(r)$ is known to smoothly vanish towards the core centre on the scale of the coherence length $\xi$. The cores host quantized quasiparticle energy levels known as Caroli-de Gennes-Matricon (CdGM) bound states [Caroli {\it et al.,} Phys. Lett. v. 9, 307 (1964)]. In pure materials, the spectrum of the low-lying CdGM states has the characteristic level spacing $\sim \Delta_0^2/E_F$, where $E_F$ is the Fermi energy and $\Delta_0$ is the bulk gap. In disordered ones, the CdGM states shift and broaden due to scattering. Here, we show, both experimentally and theoretically, that the situation is completely different in granular Nb films, which are commonly used in superconducting electronics. In these films, in which the  grains are smaller than $\xi$, the gap $\Delta$ in the quasiparticle spectrum reduces towards the vortex core centres by discrete jumps at the grain boundaries. The bound states adapt to the local environment and appear at unexpectedly high energies. Both $\Delta(r)$ and bound states form a puzzle-like spatial structure of the core, elements of which are whole grains. Our discovery shakes up the established understanding of the quantum vortex and encourages a reconsideration of the vortex motion and pinning mechanisms in granular superconductors.} 











\maketitle


\bigskip


\begin{figure*}[h]
\centering  
\includegraphics[width=0.99\textwidth]{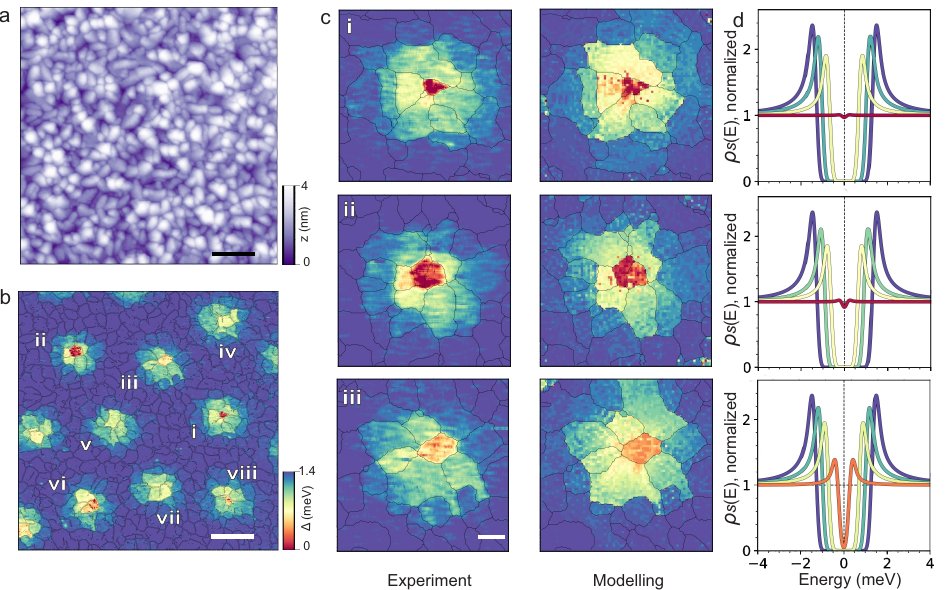}
\caption{\textcolor{black}{\textbf{Vortex cores in granular niobium.}} \textbf{a} --  300 nm $\times$ 300 nm topographic STM image of studied Nb film showing its granular structure, with the grain size $3-10$ nm. The film roughness is about 1.5 nm R.M.S.  \textbf{b} -- Color-coded spatial map of the superconducting gap (gap map) of the sample in the region shown in (\textbf{a}) at $T=1.1$ K after field-cooling the sample at 0.25 T. The grain boundaries are outlined by thin black lines. The gaps evolve a little inside each grain, but vary strongly across the grain boundaries. White labels identify eight vortices studied. \textbf{c} -- Left column: detailed gap map of the vortex cores 'i', 'ii' and 'iii'. Right column: respective gap maps calculated within the Bogoliubov-de Gennes framework (\textcolor{black}{see Methods and Section 3 of Supplementary Information}). Scale bars in (a-b) correspond to 50 nm, and in (c) -- to 10 nm. \textbf{d} -- Density of quasiparticle excitations $\rho_s(E)$ found in different grains. The colors of curves correspond to those of grains in (c), left column. The dependencies $\rho_s(E)$ are extracted from the fits of the experimental STS data (\textcolor{black}{see Methods and Section 2 in Supplementary Information}). In the centre of the vortex core 'iii' the gap does not vanish. The same holds for cores 'iv', 'v' and 'vii'. 
}
\label{Fig1}
\end{figure*}

The  dynamics of the quantum vortex lattice defines essential characteristics of superconductors -- critical fields and currents -- and thus conditions their applications in energy transport, superconducting electronics, and quantum technology~\cite{Gorkov1975}. 
Uncovering the microscopic mechanisms of the vortex motion and pinning\cite{larkin1979pinning,embon2015probing} is therefore crucial, stimulating the development of novel experimental approaches probing them in superconducting materials and devices on the local scale   \cite{karapetrov2012evidence,anahory2014three,embon2017imaging,larionov2025peculiarities,Aladyshkin2025,2025_Hovhannisyan,stolyarov2022revealing,kalashnikov2024demonstration,hovhannisyan2021lateral}.

The theoretical prediction of vortices in type-II superconductors by Abrikosov\cite{abrikosov1957magnitnykh} also marked the beginning of a new era in fundamental research, establishing these topological defects as essential quantum objects of study\cite{blatter1994vortices,brandt1995flux}. Each vortex contains a line-shaped singularity around which the phase of the quantum condensate increases exactly by $2\pi$. At the singularity, the superconducting gap vanishes, forming a potential quantum well for the Andreev quasiparticles which interfere inside the vortex core and form a set of discrete CdGM bound states\textcolor{red}{\cite{caroli1964bound}}. 
By probing the local quasiparticle density of states (LDOS), the scanning tunneling microscopy and spectroscopy (STM/STS) enabled the direct visualization of both vortex cores and CdGM states in clean superconductors, where the mean free path $\ell$ is much longer than the coherence length $\xi$ \cite{kramer1974core,hess1989scanning,menard2015coherent,bardeen1969structure}.
Though in dirty materials, where $\ell$ is much shorter than $\xi$, the CdGM states were not observed~\cite{Cren2011,stolyarov2014ex,stolyarov2018expansion}. 
In fact, the conventional (potential) quasiparticle scattering modifies the structure of sub-gap bound states, with the wave functions localized inside the vortex cores. This modification affects both the energy spectra of quasiparticles, including the residual spectral mini-gap, and the spatial structure of the electron-hole wave functions, which shape the LDOS probed by STM/STS. A strong homogeneous disorder, $\ell \ll \xi$, suppresses the zero-bias anomaly in the vortex centre and completely smears out all spectral features related to the discreteness of the CdGM levels\textcolor{red}{\cite{de2018superconductivity,gygi1991self,golubov1994,klein1990density,pottinger1993large}}. This results in a featureless quasiparticle spectrum that mimics the normal-state LDOS. 

On the other side, the quasiparticle scattering on extended individual objects, such as columnar and linear defects, twin boundaries, or surface defects, can result in an increase of the mini-gap in the spectrum of quasiparticles bound in the core, and be responsible for the vortex pinning and changes in the LDOS~\cite{Melnikov2009,Samokhvalov2009,Melnikov2020t,Khodaeva2022}. 
The physics of these effects can be elucidated on the basis of a standard one–dimensional Andreev problem with the gap function experiencing the $\pi$-shift along the quasi-classical trajectories crossing the vortex core \cite{volovik1993vortex,volovik2003universe}. The solution of such problem provides an exact zero-energy level which can be shifted to finite energies by any trajectory kink due to the change in the corresponding superconducting phase difference between the trajectory ends. 

Though, in typical granular superconductors, as in sputtered Nb films studied in the present work, none of the above considerations hold. There, Nb grains connect to each other and form a closely packed superconducting network (Fig.~\ref{Fig1}a). 
The mean free path is limited mainly by the grain size,  $\ell=3-10$ nm, which is only a factor of 4-12 smaller than the coherence length $\xi_{Nb}=38$ nm of pure Nb. The well-known expression for the effective coherence length $\xi$ in the dirty limit gives $\xi\simeq\sqrt{\xi_{Nb}\ell}=10-20$ nm, that matches well the experimentally determined value $\xi=12$ nm. Therefore, the system could be seen as a nano-network composed of clean metallic grains of size $\sim\ell\lesssim \xi$, linked together via disordered barriers. Such a landscape is far from being either a homogeneously disordered medium or a collection of rare individual defects. Consequently, $\Delta(r)$ and the CdGM states should adjust to this environment, segmented into discrete, confining grains. The experimental data and further analysis reveal that the evolution of both parameters arises from an interplay between the geometric confinement imposed by the granular structure and the spatial profile of the quasiparticle wave functions.




\section*{Results}
\

The STM/STS experiments were performed at 1.1 K on as-grown magnetron-sputtered Nb films (\textcolor{black}{see Section Methods and Section 1 of Supplementary Information}). To avoid sample contamination and modification of the electronic states at the surface, the films were transferred to the STM chamber without breaking ultrahigh vacuum. Despite the granular structure, the STS acquired in a zero-magnetic field revealed the tunneling conductance to be spatially homogeneous, presenting a gap $\Delta_0 = 1.35$ meV, slightly below the superconducting gap of bulk Nb.


When an external magnetic field is applied perpendicularly to the film plane, the tunneling conductance becomes spatially inhomogeneous, presenting strong variations of the gap. The quasiparticle gap map $\Delta(x,y)$ of the field-cooled (at 0.25 T) sample is presented in Fig.~\ref{Fig1}b (\textcolor{black}{the raw STS data and the gap map calculations are detailed in Section 2 of the Supplementary Information}). The gap map clearly demonstrates the presence of a disordered Abrikosov vortex lattice with vortex cores appearing as light spots (smaller gap).
The gap $\Delta \simeq\Delta_0$ is still observed in the LDOS between vortices (blue regions in Fig.~\ref{Fig1}b, blue curves in Fig.~\ref{Fig1}d).

The gap map in Fig.~\ref{Fig1}b demonstrates a large variety of vortex core shapes, as well as different $\Delta(x,y)$  profiles near the core centres. This is at variance with identical cores measured in homogeneous superconductors. Furthermore, the STS data revealed several unusual characteristics of the vortex cores, which are at odds with all previously reported results: 1 -- the shape of the cores is not round but irregular; 2 -- $\Delta(x,y)$ evolves by jumps, and its spatial evolution is specific to each vortex; 3 -- the jumps occur at the grain boundaries; 4 -- within a given grain, the gap remains constant (for small grains) or slightly varies (for large ones); 5 -- in the core centres, the gap is reduced but does not vanish. It becomes clear that $\Delta(x,y)$ in the vortex cores is strongly affected by the local granular structure of the sample. 

Building on these observations, we performed a more detailed analysis of cores. Three selected vortex cores are presented in Fig.~\ref{Fig1}c. Each core exhibits unique spectral evolution $\Delta(x,y)$, due to its location within a specific  granular environment. The vortex cores 'i' and 'ii' demonstrate nearly complete gap suppression in their central grain (dark red regions in Fig.~\ref{fig:sing}c, 
dark red curves in Fig.~\ref{fig:sing}d). A full gap suppression is indeed expected at the location of $2\pi$-singularity; it was experimentally observed in the vortex cores of homogeneous type-II superconductors~\cite{Ning2010,Cren2011}. Though the vortex core "iii" shows only a very partial reduction of the gap in the LDOS; the residual gap of a finite width $\simeq0.4$ meV remains clearly observable in the core centre (orange region and orange curve in Figs.~\ref{fig:sing}c,d, respectively); a location in which the gap would vanish is absent. The same holds for cores ’iv’, ’v’ and ’vii’, in the centres of which a residual gap of about $0.2-0.8$ meV exists. Clearly, the revealed variety of vortex core configurations directly reflects the granular environment, where the microscopic structure of each vortex core as well as the location of $2\pi$-singularity are determined by the local granular landscape.

\section*{Numerical Reconstruction}
\
\begin{figure}[h]
\includegraphics[width=0.5\textwidth]{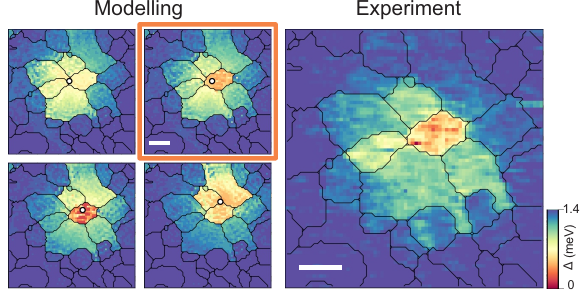}
\caption{\textcolor{black}{\textbf{Locating the 2$\pi$-singularities in the vortex cores.}} Left column: four examples of the gap map calculated for the vortex core 'iii' (see Fig.~\ref{Fig1}) considering different locations for the 2$\pi$ singularity (shown by white circles). The gap map encountered in red gives the best match with the experimental gap map. Right  column: experimental gap map of the core 'iii'. White scale bars correspond to 10 nm. }

\label{fig:sing}
\end{figure}

The observed complex spatial evolution of the superconducting gap in the tunneling conductance spectra (Fig.~\ref{Fig1}b) was successfully reproduced in numerical simulations within the Bogoliubov-de Gennes approach (\textcolor{black}{see Section 3 of Supplementary Information}). We considered that the grain size/height variations do not have a significant effect on the quasiparticle quantum mechanics compared to the presence of grain boundaries. Therefore, to construct the effective Hamiltonian, the sample was considered as a two-dimensional system of grains separated by atomically-thin boundaries. The exact position of the grain boundaries was determined from the experimental topographic data presented in Fig.~\ref{Fig1}a (see Methods and Section 2 of Supplementary Information). Then, for each studied region, a dense grid was superimposed onto the obtained two-dimensional granular landscape. To account for lower inter-grain transmission compared to the intra-granular one, the electron tunneling amplitude was fixed to $t_0$ for neighboring sites of the grid belonging to the same Nb-grain; for neighboring sites belonging to different grains, it was set to $t_1$, with the ratio $t_1/t_0<1$ used as a parameter to adjust (in practice, in a rather narrow window $0.1-0.5$ delimited by well-established properties of sputtered Nb-films; \textcolor{black}{for details, see Section 3 in Supplementary Information}). 


Remarkably, even with these rough simplifications, the simulations presented in right column of Fig.~\ref{Fig1}c nicely reproduce the results of STM/STS measurements (left column). Minor quantitative discrepancies can be attributed to the neglected variations of the inter-granular coupling $t_1$ from one junction to another, to a spatially non-uniform pairing strength, to disorder effects, and, last but not least, to the three- and not two-dimensional granular structure of the real sample. Crucially, these factors do not alter the physics of the gap evolution, which is properly captured by theory.

Furthermore, our numerical calculations enable locating the positions of $2\pi$-phase singularities, something that STM/STS measurements cannot reveal. In Fig.~\ref{fig:sing}, right column, we recall the experimental gap map pattern observed in the vortex core "ii". In the left column, we present the results of four $\Delta(x,y)$ calculations performed for the same sample region while keeping all parameters fixed, but placing the vortex singularity in slightly different locations (the positions of the singularity are marked by small white circles). It is straightforward to see that tiny variations in the position of the singularity lead to strong variations in the resulting $\Delta(x,y)$ pattern.

\section*{Discussion} 
\

\begin{figure}[h]
\centering  
\includegraphics[width=0.49\textwidth]{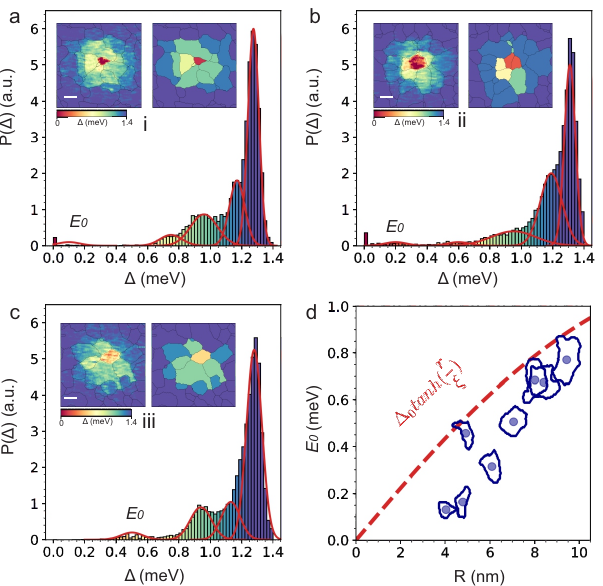}
\caption{ \textcolor{black}{\textbf{Identification of the bound states.}}  \textbf{a--c} -- Histograms depicting the energy distribution of the gap within regions where vortices are localized. \textbf{a}: vortex i; \textbf{b}: vortex ii; \textbf{c}: vortex iii. The lowest-lying states are indicated by $E_0$. Insets compare the experimentally observed gap distribution (left) with that predicted by the puzzle-based model of CdGM states in a granular superconducting film (right). \textbf{d} -- Variation of $E_0$ as a function of size of the central grain containing the 2$\pi$- singularity. Red dashed line: conventional variation of the order parameter in the core of a homogeneous superconductor with $\xi$ = 12 nm and $\Delta_s$ = 1.35 meV.
} 
\label{Fig3}
\end{figure}

The grain boundaries, being weak regions of the landscape, would logically host $2\pi$-singularities. Such a situation, predicted by A. Gurevich~\cite{Gurevich1992}, indeed occurs. In the vortex 'ii', for example (Fig.~\ref{Fig1}c), the singularity sits between two small grains, forming the vortex centre. Though this is not the case for vortices 'iii', 'iv', 'v', and 'vii'. The top-left gap map in Fig.~\ref{fig:sing} displays the calculated gap map for the vortex core 'iii' when the 2$\pi$-singularity is placed at the grain boundary. Such a configuration leads to poor correspondence between experimental and numerically generated $\Delta(x,y)$ patterns. Rather, an almost perfect agreement is reached when the singularity is placed close to, but not exactly at the grain boundary. In general, the calculations demonstrate a strong tendency for the vortex cores to nucleate near grain boundaries, but not always at them.\footnote{It should be noted that the STM/STS probes the vortex structure at the topmost granular layer and that the calculations reproduce the cores assuming a two-dimensional granular landscape. However, the studied films are thicker than the lateral grain size by a factor of 3-8 and may possess a more developed granular structure. To reduce the total vortex core energy, the 2$\pi$-singularity can 'optimize' its twisted path when crossing the sample's bulk to emerge at the surface off the grain boundary. The trajectory of the singularity is affected by interactions (pinning) with grain boundaries, as well as with other vortices in the lattice and with individual magnetic defects typical in technical-grade niobium (see Section Methods). Such magnetic defects host specific bound states~\cite{menard2015coherent}; they appear in the gap maps as tiny bright spots  (in Fig.~\ref{Fig1}b, few of them are visible in the inter-vortex region). }

\begin{figure*}[h]
\centering  
\includegraphics[width=0.99\textwidth]{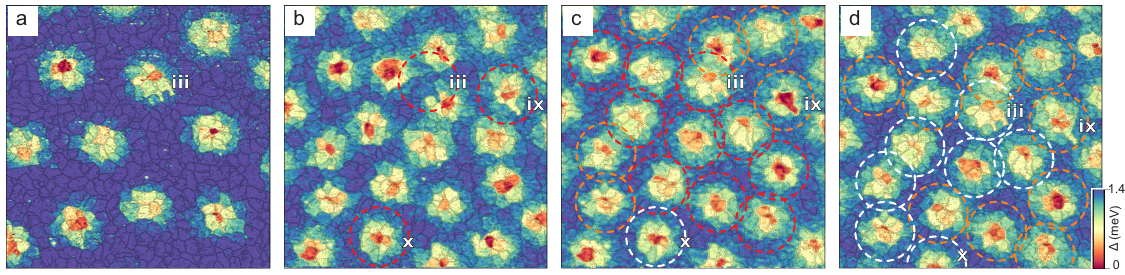}
\caption{ 
\textcolor{black}{ \textbf{Ultra-slow vortex dynamics.}}  \textbf{a} -- Vortex lattice after field-cooling the sample at 0.25 T. In total, 10-11 vortices are visible in this 300 nm × 300 nm area. \textbf{b} -- Vortex lattice in the same region after increasing the applied filed to 0.5 T and a 6 hours delay. The number of vortices increased to 20-22; their locations are completely different from (a). Red circle 'iii' shows the former position of the vortex 'iii'; the circles 'ix' and 'x' are examples of new vortices. \textbf{c} -- The same as in (b) but after the time delay of 12 hours after the field increase. With respect to (b), the number of vortices is the same but their positions and shapes changed significantly. Most of vortices moved by distances $\sim \xi$ (red circles), a few by distances $\ll \xi$ (orange circles); only one vortex, marked by 'x', did not move (white circle). \textbf{d} -- The same as in (c) but after the time delay of 18 hours after the field increase. With respect to (c), the vortex displacement is almost zero; only some vortices moved on a very local scale (orange circles), while the positions of others remained unchanged (white circles). 
} 
\label{Fig4}
\end{figure*}



For all studied cases, both experiment and calculations agree that the excitation gap $\Delta$ does not necessarily vanish at the singularity even though the pair potential $\Delta_s$ certainly does. This effect, observed in several STM/STS studies~\cite{Samokhvalov2009,Melnikov2020t,Kiselov2023,Khodaeva2022,Chen2024,maggio1995direct,pan2000stm}, can be qualitatively explained by the disruption of perfect quantum interference. In homogeneous (on the scale of $\xi$) superconductors, the exact gap vanishing results from a perfect destructive interference of all components of the Cooper pair wave function, whose phases change by $\pi$ upon crossing the singularity. However, in inhomogeneous (e.g., granular) systems, scattering at grain boundaries introduces kinks in quasiparticle trajectories, altering the amplitudes and phases of waves arriving at the singularity. This prevents perfect cancellation, spatially shifting the point of minimal amplitude away from the singularity itself. Consequently, no single location exists where all components of the many-body wave function cancel exactly, allowing a finite mini-gap to persist in the vortex centre.
Theoretically, this is supported by the one-dimensional Andreev problem, where an exact zero-energy level is shifted to finite energies by any trajectory kink that changes the superconducting phase difference. This mechanism explains how the mini-gap can open and close as the vortex position shifts relative to the scatterers. Furthermore, in systems with complex interfering trajectories, standing quasiparticle waves can form, leading to unusual local density of states (LDOS) profiles~\cite{Hou2025,hayashi1998low}. 
Thus, both the behaviour of the mini-gap and the resulting LDOS profiles provide a unique fingerprint of the sample disorder, offering a crucial key to understanding the non-trivial interplay between conventional and Andreev scattering processes. In this, our findings confirm existing theoretical predictions \cite{hayashi1996star,hayashi1997effects,larkin1998,koulakov1999}.

A good agreement between experimentally and theoretically obtained $\Delta(x,y)$ maps warrants that the theory also captures the evolution of the quasiparticle bound states caused by the granular structure of the sample. We found that the grain boundaries cause large shifts of the bound states to higher energies and lead to energy level bunching. The results are summarized in Fig.~\ref{Fig3}. The energy level histograms for vortex cores 'i', 'ii', and 'iii' are presented in Fig.~\ref{Fig3}a,b, and c, respectively. The up-shift of the energy levels is stronger for vortices in which the singularity is located at or close to larger grains. There, the lowest energy bunch $E_0$ can even approach the gap edge (Fig.~\ref{Fig3}d). This is the case, for instance, of the core 'v', $E_0\simeq 0.8 \Delta_0$, which is several orders of magnitude higher than the energy of conventional CdGM low-lying levels, $E_0\sim \Delta_0^2/E_F \ll \Delta_0$. Such a strong up-shift can be explained by the fact that in granular materials, the singularity sits at or close to the grain boundary. Due to this, on the scale of $\xi$, the gap in the quasiparticle excitation spectrum is present quite everywhere inside the vortex core; the bound states form above this remaining gap. This is in contrast with homogeneous superconductors in which the gap linearly vanishes towards the singularity, resulting in low lying CdGM levels.

The color code in Fig.~\ref{Fig3}a-c points out another remarkable feature: A correlation between the energy level bunches and the spatial variations of the quasiparticle LDOS. Using the extracted energy levels as fingerprints, we can identify specific grains whose electronic states most closely match these energies. The resulting reconstruction (shown in the insets) reveals the vortex core composite structure as an arrangement of participating grains. The correlation is so clear that one could naively suggest each energy bunch is a set of discrete levels representing quasiparticle bound states forming independently in each grain. Indeed, each grain is surrounded by grain boundaries that act as barriers for quasiparticles. If the barriers are infinite, the quasiparticles are fully backscattered and form bound states inside each grain separately (see Section 3 of Supplementary Information). Though, this picture is oversimplified, since the experimentally observed vortex lattice (Fig.~\ref{Fig1}b) is the direct proof of the macroscopic phase coherence that extends far over grains. This experimental finding also proves the existence of 2$\pi$-phase accumulation around each vortex core. Also, the expected reduction of the gap towards the core centres is globally respected for all studied vortices. Therefore, a more realistic picture is that the quasiparticle bound states form owing to two main contributions: (i) -- the overall vortex core pair-potential that vanishes towards the singularity and has 2$\pi$-phase winding (exactly as considered in the original CdGM work) and, (ii) -- additional barriers at the grain boundaries leading to confinement in individual grains. 
The puzzle-like spatial organization of states emerges naturally from this picture, showing how the vortex core states are constructed by the interplay between the local granular structure of the film and the spatial profile of the total potential for quasiparticles.
Note that the reorganization of core states under strong disorder was theoretically anticipated a long time ago~\cite{gygi1991self, volovik1993}.

Fig.~\ref{Fig3}d represents, for eight analyzed vortices, the energy position of the lowest energy bunch $E_0$ as a function of the size of the central grain. It is clear that, on average, the data points follow the rising evolution of the order parameter; yet, they all appear below it. This effect can be related to the granular structure of the film~\footnote{Indeed, in the vortex cores of homogeneous superconductors, the gap smoothly varies with the distance $r$ from the singularity, as $\Delta(r) = \Delta_0 \tanh{\left( r/\xi\right)}$ (red curve in Fig.~\ref{Fig3}d). However, in the granular case, the quasiparticle gap inside the central grain of a radius $R<\xi$ remains constant (because of multiple scattering from the grain boundaries at $R$); its grain-averaged value is 

\begin{eqnarray}
\langle{\Delta}\rangle \sim \frac{1}{\pi R^2} \int\limits_{0}^{R} \Delta_0 \tanh\left(\frac{r}{\xi}\right)\,2\pi r\,dr\simeq \frac{2}{3} \Delta(R)< \Delta(R).
\end{eqnarray} For the smallest grains, $\langle{\Delta}\rangle$ should be even lower because of a stronger effect of the 2$\pi$-singularity on the average gap.  }.

The peculiar vortex core structure influences the pinning and dynamics of the vortex motion. In Fig.~\ref{Fig4} we present a series of the gap maps acquired in the same region of the sample. The first one was obtained after field-cooling the sample at 0.25 T; it is identical to that presented in Fig.~\ref{Fig1}b. On this map, approximately 10-11 vortices are visible. After the acquisition of these STS data, the applied field was increased to 0.5 T. The gap maps (b), (c), and (d) present the vortex arrangements observed, respectively,  6, 12, and 18 hours after the field increase. These maps show two major effects. The first one is an increase of the vortex density, between (a) and (b), by the expected factor of two, caused by the doubling of the field intensity. Importantly, all vortices appearing in (b) occupy new locations compared to (a), demonstrating a rather low overall pinning in the film. Though some locations play the role of moderate pinning centers—those marked by 'iii",'ix', and 'x' --when occupied, host vortex cores exactly in the same positions, as evidenced by their peculiar fingerprints. The second effect is that the positions of the vortex cores continue to vary for many hours after the field has increased. This ultra-slow vortex dynamics can be appreciated by following the colored circles positioned in the same locations in (b), (c), and (d). On these maps, red, orange, and white circles indicate, respectively, vortices that moved significantly ($\gtrsim \xi$), a few ($\ll\xi$), with respect to previous maps, or remained completely pinned. While in (c) most of the circles are red, indicating significant vortex motion over the first 6-12 hours, they are orange or white in (d), thus demonstrating the vortex slowing down over time. Consequently, there are at least two regimes in the vortex motion: the well-known avalanche-like one, which occurs at short time scales after the field increase, specifically between (a) and (b) This fast dynamics is inaccessible in a STM/STS experiment, while a very slow, viscous-like motion -- a vortex glass -- is observed over hours. Therefore, the vortex lattice in granular superconductors can be viewed as a fluid in which several regimes of motion and time scales are involved. 

Note that such tiny vortex displacements would be difficult to detect in homogeneous superconductors, while the presence of the granular landscape in our case makes determining the precise vortex position an easy task, despite the lateral drift of the piezo-scanner of the STM unit. Furthermore, a high sensibility of the spatial gap distribution to the exact location of the singularity (Fig.~\ref{fig:sing}) helps in appreciating the vortex motion at yet shorter scales. 

\section*{Conclusions} 
\

In summary, we discovered that in the vortex cores of granular niobium films, the gap in quasiparticle excitation spectra evolves by jumps at the grain boundaries, whereas outside the cores, it is spatially homogeneous. The core appears in scanning tunneling spectroscopy experiments and in numerical calculations as a coherent quantum puzzle mapping the local granular landscape. The topological 2$\pi$-singularities are not obviously located at the grain boundaries, and depending on their exact positions, the gap $\Delta$ in the spectrum of elementary excitations is still observed in the vortex core centres. We also discovered a very slow regime of vortex motion that we were able to track 
over hours after the initial perturbation of the vortex lattice. This regime, related to the granular structure of the system, resembles the viscous motion of glass. Our findings bridge the gap between the widely-studied CdGM states in clean systems and the theoretical understanding of vortex core structure in disordered environments.

\section*{Methods} 
\

Thin Nb films were fabricated using standard DC-magnetron sputtering of an Nb target (purity 99.95\%) in Ar plasma. The target contains up to 0.003\% residual magnetic impurities, such as Fe and Ni.
The initial pressure in the growth chamber was $p_0=5\times 10^{-9}$~mbar. During growth, Ar pressure was $p_{\text{Ar}} = 4\times 10^{-3}$~mbar, the power was $P_{\text{DC}} = 200$~W, and the voltage was $V_{\text{DC}} = 238$~V. At these conditions, the deposition rate was 0.13 nm/s, resulting in a formation of 25-nm-thick granular Nb films. After the sputtering, the pressure in the growth chamber was reduced back to $p_0$, and the samples were transferred under ultrahigh vacuum conditions directly to the STM/STS unit (Tyto$^{\text{TM}}$ LT-STM/AFM by SPECS 0-3 T UHV system). Then, the transport properties of the samples were examined. To do this, the films were extracted from the STM chamber and exposed to air, thus unavoidably contaminating and oxidizing their surfaces. Though even contaminated films showed a complete resistive transition to the superconducting state at $T_c=7.3$\,K.


Our STM/STS experiments followed well-established protocols for probing vortex core states \cite{hess1989scanning,hess1991stm,hoogenboom2000shape,kohen2006probing,guillamon2008,fischer2007scanning}. The experiments were conducted in ultrahigh vacuum (base pressure below $1\times 10^{-10}$~mbar) at a temperature of 1.1 K. To enhance the spectral resolution, Prior to STM/STS measurements, tungsten STM tip was functionalized by picking up a grain of Nb and stabilizing it on the tip apex. This results in a superconducting tip and better spectral resolution in the STS measurement, yet it requires an additional procedure to extract the excitation spectra of the sample. The topographic STM data (Fig.~\ref{Fig1}a) were acquired in the constant current mode (tunneling current $I=$ 60 pA, sample bias $V=$ 10 mV). The STS measurements were performed by freezing the position of the tip above the sample, by acquiring local $I(V)(x,y)$ characteristics, and by repeating this procedure in every location of the region of interest.

The shape $\rho^{\,}_s(E)$ of the tunneling DOS of the sample presented in Fig.~\ref{Fig1}d, along with the gap maps  $\Delta(x,y)$ presented in Figs.~\ref{Fig1}-\ref{Fig3},
were numerically derived from raw $I(V)(x,y)$ local tunneling spectra following the procedure detailed in Section 2 of Supplementary Information. While $dI/dV(V)(x,y)$ tunneling data already show the puzzle-like spatial variations of the tunneling spectra in the vortex cores (see Supplementary Fig.1b), the knowledge of $\Delta(x,y)$ enables a direct comparison with the results of numerical simulations.

The modeling of the vortex core states was realized within the Bogoliubov-de Gennes framework; the calculations were performed on a dense two-dimensional grid mapping the granular landscape of the real sample (see Section 3 of Supplementary Information). The landscape was extracted directly from topographic STM images using the well-known difference-of-Gaussian approach\cite{Marr-80,Aladyshkin-JPCC-24}. 



\hspace{1cm}

\bibliography{Biblio}

\bigskip

\section*{Acknowledgements} \hspace{1cm}

This work is supported by a grant from the Ministry of Science and Higher Education of the Russian Federation No 075-15-2025-010, dated 28.02.2025.
V.S.S. and D.R. thank the 'Metchnikov mobility program' of the French Embassy in the Russian Federation and the CRYSTOP French-Russian ANR project.  

\section*{Author contributions} \hspace{1cm}

V.S.S. and D.R. initiated and supervised the project. V.S.S. built the experimental setup, V.S.S., D.I.K., O.V.S., and D.B. elaborated Nb films, V.S.S., D.I.K., and O.V.S. performed STM/STS experiments; V.S.S., D.P., A.Yu.A. and V.N. extracted relevant information from the experimental data; V.S.S., V.N., A.V.K., A.A.Sh. and A.V. made numerical calculations; all coauthors contributed to the data analysis;  V.S.S., V.N., A.S.M., A.A.G., M.Yu.K., T.C., A.V. and  D.R. proposed the explanation of the observed phenomena; V.S.S., V.N., A.V.K., A.S.M., A.Yu.A. and D.R. wrote the draft of the manuscript, reviewed it, and edited its final version with essential contributions from the other authors.  

\hspace{1cm}
\subsection{Data availability.}
Authors can confirm that all relevant data is included in the paper and its supplementary information files. Additional data are available on request from the authors.

\section{Additional information}

\subsection{Supplementary Information} accompanies this paper at https://doi.org/.....
\subsection{Competing interests:} The authors declare no competing interests.
\subsection{Reprints and permission} information is available online at http://npg.nature.com/
reprintsandpermissions/
\subsection{Publisher's note:} Springer Nature remains neutral with regard to jurisdictional claims in published maps and institutional affiliations.

\end{document}